\documentclass[preprint,showpacs,preprintnumbers,amsmath,amssymb]{revtex4}
\usepackage{epsfig}
\usepackage{euscript}
\DeclareFontFamily{OT1}{rsfs}{}
\DeclareFontShape{OT1}{rsfs}{m}{n}{<5> rsfs5 <7> rsfs7 <10> rsfs10}{}
\DeclareSymbolFont{mathrsfs}{OT1}{rsfs}{m}{n}
\DeclareSymbolFontAlphabet{\mathrsfs}{mathrsfs}
\newcommand{\R}{\mathrsfs{R}}
\newcommand{\p}{\mathrsfs{P}}

\newcommand*{\Cet}[1]{|{#1}\rangle}
\newcommand*{\Bra}[1]{\langle{#1}}

\newcommand{\Up}{\uparrow}
\newcommand{\Dw}{\downarrow}

\newcommand{\UpCet}{\Cet{\Up}}
\newcommand{\DwCet}{\Cet{\Dw}}
\newcommand{\UpUp}{\Bra{\Up}\Cet{\Up}}

\newtheorem{axiom}{Axiom}

\begin{document}

\title{Applications of  The Information Model of The Collapse Phenomena: The Mathematical Model of Everett's Worlds in The Case of The Measurement of A Spin $1/2$ Projection}

\author{Denys Bondar}
 \altaffiliation{Department of Physics and Astronomy, University of Waterloo, 200 University Avenue West, N2L 3G1 Waterloo, Ontario, Canada.}
 \email{dbondar@scimail.uwaterloo.ca}

\date{\today}

\begin{abstract}
The information model of the collapse phenomena is further advanced.  We discover an important property of the model - the death point effect. The $\p$ function approach  is presented to construct the manifest form of the function of risk. We clarify a close connection of the model with the Extended Everett Concept. The model is also reformulated as an automaton. Examples are considered.
\end{abstract}


\maketitle

\section{Introduction}
This paper devotes to applications of an approach that has been offered in \cite{Bondar} for describing the state collapse phenomena. In this article, this approach will be called the information model of the collapse phenomena. We shall follow the following definition of the collapse:
\begin{quotation}
``In its most basic formulation, quantum theory encodes the preparation of a system in a pure 
quantum state, a unit vector $\psi$ in a Hilbert space $\EuScript{H}$. Observables are modelled by (say, 
nondegenerate) self-adjoint operators on $\EuScript{H}$. The expectation value of an observable $A$ in 
a state $\psi$ is given by $\langle\psi, A\psi\rangle$. If $a$ is an eigenvalue of $A$ and $\psi_a$ a unit eigenvector, and 
information concerning $A$ is somehow extracted from the system, then the probability for the 
value $a$ to be observed is $|\langle\psi_a,\psi\rangle|^2$. If this observation is indeed made, then the subsequent 
behaviour of the system is predicted using the pure state $\psi_a$ as a starting point. This is called 
{\it state collapse}.'' (\cite{Janssens}, P.~9845)
\end{quotation}

The information model of the collapse  has been formulated based on ideas from \cite{Brillouin1, Neumann, Menskii, Mensky}. Treating the state collapse as an information process, this model allows us to include an observer in the description of a quantum mechanical measurement. Moreover, according to this model, the observer plays a crucial role in the collapse phenomena. In the current paper we will further develop the information model of the collapse. A close connection between the model and Everett's interpretation of quantum mechanics \cite{Menskii, Mensky} will be clarified. A method of the construction of the manifest form of the function of risk will also be considered. We will show that the information  model can be reformulated as an automaton \cite{Gecseg}. Generally, we will see that ideas from \cite{Menskii, Mensky} can be mathematically defined.

The structure of the article is the following. In the first section, a review of the information model of the collapse is made. Some aspects of numerical computations in this model are discussed in the second section. The death point effect, a specific feature of the model, is clarified in the forth section. In section five, Everett's interpretation of quantum mechanics from the point of view of our model is considered. Finally, conclusions are made in the last section.

\section{Review of the information model of the collapse}

Before applying the information model of the collapse phenomena, we are going to make a brief review of this model. Let us draw the following situation. An observer wants to measure a projection of the spin of an electron on an axis 
\begin{equation}\label{NVec}
\overrightarrow{n} = (\sin\theta\cos\varphi,
\sin\theta\sin\varphi, \cos\theta), \quad 0\leq\theta\leq\pi,
\quad 0\leq\varphi < 2\pi,
\end{equation}
where $\theta$ and $\varphi$ are spherical angles. In quantum mechanics, a projection of the spin on this axis is represented by a hermitian operator
$$
\sigma(\theta,\varphi) =  \left(
\begin{array}{cc}
\cos\theta & \sin\theta e^{-i\varphi} \\
\sin\theta e^{i\varphi} & -\cos\theta
\end{array}
\right).
$$
The eigenvalues and the orthonormal eigenvectors of the operator
$\sigma(\theta,\varphi) $ are the following
\begin{eqnarray}
\sigma_n\UpCet = +1\UpCet, && \sigma_n\DwCet = -1\DwCet, \nonumber\\
\UpCet = \left(
\begin{array}{c}
\cos(\theta/2) e^{-i\varphi}  \\
\sin(\theta/2)
\end{array}\right), \quad&&
\DwCet = \left(
\begin{array}{c}
-\sin(\theta/2)e^{-i\varphi} \label{EginF}\\
\cos(\theta/2)
\end{array}\right).
\end{eqnarray}
We are interested in answering the question: ``What is the final state of the electron after the measurement if the observer wants to measure the projection of the spin on an axis given by the  angles $\theta_i, \varphi_i$ and the initial state of the electron before the measurement is $\Cet{\Psi}$?'' 

\begin{figure}
\begin{center}
\includegraphics[scale=0.3]{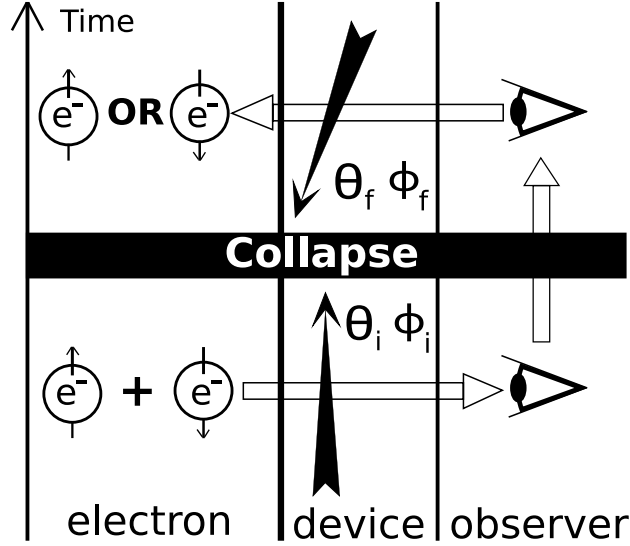}
\end{center}
\caption{The scheme of the collapse of the wavefunction of an electron at the measurement of a spin projection}
\label{fig1}
\end{figure}

According to the information model, the collapse is an information process which  is picturized on Fig. \ref{fig1}.  There are three important parts: the electron (the observed object), the device (the tool for observing, or the tool that makes the observer be able to observe) and the observer (a human being who is interested in obtaining knowledge about the electron). According to the model, there are three steps of the collapse. In the first step, decoherence, embedding the electron into the device, occurs. In the second step, having obtained information about the quantum system (by means of decoherence \cite{Janssens}), the observer changes his state and chooses what projection of the spin he wants to see after the collapse. According to this decision, the collapse occurs in the last step. About further details see \cite{Bondar}.

The mathematical formulation of this scheme can be achieved by means of three axioms. First of all, we have to mention that a state of the observer is given by the operator of the projection which he wants to measure, i. e. the state of the observer is fixed by spherical angles $\theta$ and $\varphi$. To formulate the axioms, we need to introduce the state of the observer after the collapse by $\theta_f$ and $\varphi_f$. Hereinafter, the vectors $\UpCet_i$ and $\DwCet_i$ are the eigenvectors of the operator $\sigma(\theta_i,\varphi_i) $; the vectors $\UpCet_f$ and $\DwCet_f$ correspond to $\sigma(\theta_f,\varphi_f) $.

Let us introduce a function
\begin{equation}\label{f}
f(p) = -p\ln p - (1-p) \ln(1-p).
\end{equation}
All entropies, that take place because of decoherence, can be written in terms of this function
\begin{equation}\label{fentropies}
S_i = f\left( \Big|{}_i\Bra{\Up}\Cet{\Psi} \Big|^2 \right), \qquad
S_f = f\left( \Big| {}_f\Bra{\Up}\Cet{\Psi} \Big|^2 \right), \qquad
S_{\Up} = f\left( \Big| {}_f\UpUp_i \Big|^2 \right).
\end{equation}

The first axiom follows from the negentropy principle of information (\cite{Brillouin1}, P.~153) (the conservation of the sum of entropy and information)
\begin{axiom}
\begin{equation}\label{Ax1}
S_i(\theta_i, \varphi_i)=S_f(\theta_f, \varphi_f).
\end{equation}
\end{axiom}

However, there is one exception, when the observer wants to measure the projection of the spin on the same axis, that the initial state is an eigenstate of the operator of that projection, i. e. the wavefunction of the system is either $\Cet{\Psi}=\UpCet_i$ or $\Cet{\Psi}=\DwCet_i$. In this case equations (\ref{Ax1}) is a trivial equality, because  $S_i =0$. According to the postulates of quantum mechanics, the initial state is also the final state, i. e. there is no collapse in this case. Hence the state of the observer cannot be changed which implies that $\theta_f=\theta_i$ and $\varphi_f=\varphi_i$. We will exclude this trivial case in further investigations. 

From the generalized Carnot's principle (\cite{Brillouin1}, P.~153) we reach the second axiom
\begin{axiom}
\begin{equation}\label{smin}
S_{\Up}(\theta_i, \varphi_i, \theta_f, \varphi_f) \to min.
\end{equation}
\end{axiom}
Finally, the third axiom follows from the principle ``classical alternatives: prerequisite to the existence of life'' (\cite{Menskii}, P.~402)
\begin{axiom}
$$\R(\theta_f, \varphi_f; s) \to min$$.
\end{axiom}
Here $\R$ is a real function which is called the function of risk, because  its value represents the measure of the risk of an alternative. In our problem we have two alternatives: the first one is the `spin up', the second one is the `spin down.'

Let us rewrite Eq. (\ref{Ax1}) in the terms of our parameters. We can assume without the lose of  generality that a normalized vector has the form
\begin{equation}
\Cet{\Psi} = \left(\begin{array}{c} \sqrt{\rho}e^{-i\tau}\\
\sqrt{1-\rho} \end{array}\right), \qquad 0\leq\rho\leq 1, \qquad
0\leq\tau\leq2\pi.
\end{equation}
From equation (\ref{EginF}) we reach 
\begin{equation}
\Big| {}_f\Bra{\Up}\Cet{\Psi} \Big|^2 =
\rho\cos^2\frac{\theta_f}2 + (1-\rho)\sin^2\frac{\theta_f}2 +
\sqrt{\rho(1-\rho)}\sin\theta_f \cos(\varphi_f-\tau).
\end{equation}

\section{Numerical computations}

Before computing,  it is important to clarify the symmetry of our problem.  The geometrical symmetry of the problem is indicated in Eq. (\ref{NVec}): $\theta \to \theta+\pi$, $\varphi \to \varphi+2\pi$. However, according to Eq. (\ref{EginF}), we have
$$
\UpCet \Big|_{\theta\to\theta +\pi, \, \varphi\to\varphi +\pi} = \DwCet, \qquad \DwCet \Big|_{\theta\to\theta +\pi, \, \varphi\to\varphi +\pi} = \UpCet.
$$
This leads to the invariant property of an entropy: $S(\theta+\pi, \varphi+\pi)=S(\theta, \varphi)$, where $S$ can be any of $S_f$, $S_i$ and $S_{\Up}$.
So this transformation of the coordinates ($\theta\to\theta +\pi$, $\varphi\to\varphi +\pi$)  represents the physical symmetry of the system. Therefore,  we have to restrict the region of the coordinates
\begin{equation}
0\leq\theta <\pi, \qquad 0\leq\varphi <\pi.
\end{equation}

The following remark should be made here. If  $\theta = 0$ or $\pi$ then, according to the geometrical symmetry of the spherical coordinate system, the position of a point does not depend on $\varphi$. Therefore, we shall fix $\varphi=0$ in this case.

\begin{figure}
\begin{center}
\includegraphics[scale=0.3]{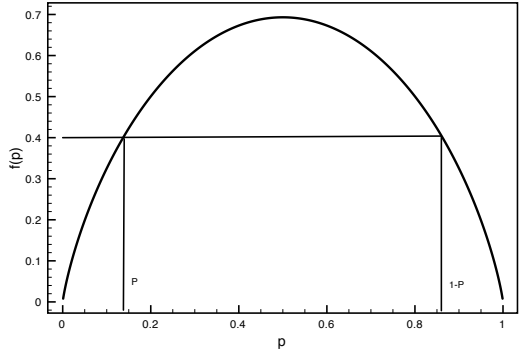}
\end{center}
\caption{The graph of the function $f(p)$}
\label{fig2}
\end{figure}

Let us come back to equations (\ref{f}) and (\ref{fentropies}). The plot of  the function $f$ is shown in Fig. \ref{fig2}. 
As it can be easily seen if $p$ ($0\leq p \leq 1$) is a solution of the equation $f(p)=S$ ($0\leq S\leq \ln 2)$, then $1-p$ is the second solution of this equation. Taking into account this property of the $f$ function and applying the property of Shannon's entropy (see for example \cite{Klir} theorem~3.2, P.~78; also \cite{Neumann} P.~385) to equation (\ref{Ax1}) and the axiom of the symmetry (\cite{Klir} axiom (S2) P.~73), we can conclude that
\begin{equation}\label{restr}
\Big| {}_f\Bra{\Up}\Cet{\Psi} \Big|^2 = \Big|{}_i\Bra{\Up}\Cet{\Psi} \Big|^2,  \qquad OR  \qquad \Big| {}_f\Bra{\Up}\Cet{\Psi} \Big|^2 = 1 - \Big|{}_i\Bra{\Up}\Cet{\Psi} \Big|^2.
\end{equation}

As far as $S_{\Up}$ is concerned, from Fig. \ref{fig2} and from the fact that $0\leq | {}_f\UpUp_i |^2 \leq 1$, we can deduce that the set of the points of the minimums of $S_{\Up}$ is equivalent to the set of the extremums of $| {}_f\UpUp_i |^2 $. The following has to be pointed out here. When we are minimizing $S_{\Up}$ the case of $S_{\Up}=0$ has to be avoided, because this case is the exception that has been mentioned after Eq. (\ref{Ax1}). Therefore, instead of the second axiom, we obtain
\begin{equation}\label{UpUpExtr}
\Big| {}_f\UpUp_i \Big|^2 \to max, \,\dfrac{}{} min \qquad AND \qquad \Big| {}_f\UpUp_i \Big|^2 \neq 0, \, 1.
\end{equation}

Finally, we reduce the problem of minimizing (\ref{smin}) with condition (\ref{Ax1}) to the problem of finding the extremums of $| {}_f\UpUp_i |^2 $ (\ref{UpUpExtr}) with restriction (\ref{restr}). This transformation is crucial not only because of the simplicity of the problem (\ref{UpUpExtr}) and (\ref{restr}) but also the last one has the most appropriate form for the numerical computations. Let us clarify this difference. Both problems are rather cumbersome to be treated analytically. Therefore, numerical methods ought to be applied. The easiest way to do this is to use special software for non linear programing \footnote{See for example: http://www-unix.mcs.anl.gov/otc/Guide/faq/nonlinear-programming-faq.html}. For example, Maple (http://www.maplesoft.com/) has the package `Optimization' which contains the procedure `NLPSolve' for computing the minimum (or maximum) of a real-valued function with constraints (about further information see Maple Help). Mathematica (http://www.wolfram.com/) for this purpose has two different functions `NMinimize' and `NMaximize'. If we try to employ these procedures to the problem (\ref{smin}) and (\ref{Ax1}), we will obtain $S_{\Up}=0$ as the minimum (It should be noticed that for solving this problem Mathematica is better than Maple). But, this result must be avoided. In addition to this, all mentioned procedures do not allow us to exclude correctly this minimum. However, the situation is completely different in the case of the problem (\ref{UpUpExtr}) and (\ref{restr}).

\begin{figure}
\begin{center}
\includegraphics[scale=0.3]{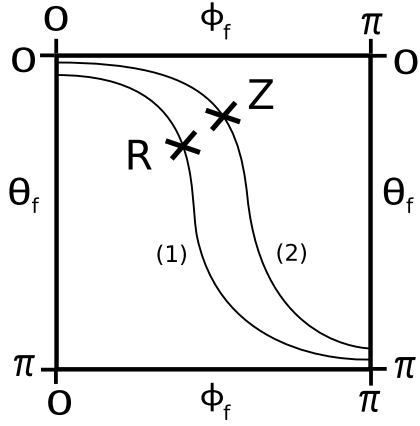}
\end{center}
\caption{The lines of the values of $\theta_f$ and $\varphi_f$  according to the first axiom for the case of  $\theta_i=\pi/4, \varphi_i=\pi/2, \rho=0.4, \tau=0$}
\label{fig3}
$Z$ ($\theta_f=0.785, \varphi_f=1.571$) is the point where $S_{\Up}=0$, \\
$R$ ($\theta_f=0.862, \varphi_f=1.197$) is the point of the minimum of $S_{\Up}=0.0980$ on the second line. 
\end{figure}

Despite the fact that the problems (\ref{smin}); (\ref{Ax1}) and  (\ref{UpUpExtr}); (\ref{restr}) are equivalent, the treating of the last one brings us correct results (for solving this task Maple is as good as Mathematica). Let us look at Fig. \ref{fig3}. There we can see the two lines which represent the set of the values of $\theta_f$ and $\varphi_f$ according to the first axiom (the case of $\theta_i=\pi/4, \varphi_i=\pi/2, \rho=0.4, \tau=0$ is considered). $Z$ is the very point ($S_{\Up}=0$) that must be ignored. But taking into account that $S_{\Up}$ is a continuous function of the variables $\theta_f$ and $\varphi_f$, {\it we must neglect not only the point $Z$ but also whole line $(2)$.} Of course, we cannot split these two lines during solving the task (\ref{smin}) and (\ref{Ax1}). However, we are able to do this by means of Eq. (\ref{restr}). In the considered case the correct result is the $R$ point.

\section{Death point effect}

Let us continue considering the example on Fig. \ref{fig3}. We have computed the values of $\theta_f$ and $\varphi_f$. Let us think that after the collapse, the electron has the $+1/2$ projection, i. e. the final wavefunction of the electron is one with $\rho= \cos^2 (\pi/8)$, $\tau=\pi/2$. We are going to compute the final state of this system after the next mesurment, i. e. we are supposing that this final state is the initial state and $\theta_i =0.862$ and $\varphi_i=1.197$.
\begin{figure}
\begin{center}
\includegraphics[scale=0.3]{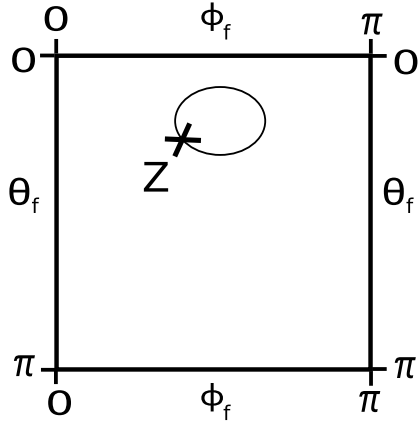}
\end{center}
\caption{The lines of the values of $\theta_f$ and $\varphi_f$  according to the first axiom for the case of  $\theta_i=0.862, \varphi_i=1.197, \rho=\cos^2 (\pi/8), \tau=\pi/2$}
\label{fig4}
$Z$ ($\theta_f=0.862, \varphi_f=1.197$) is the point where $S_{\Up}=0$.
\end{figure}
On Fig. \ref{fig4} we can see the set of the values of $\theta_f$ and $\varphi_f$ for this case. On the contrary to the previous situation, the set has the form of one line. As far as the $Z$ point is lying on this curve, as it was in the previous case, we might ignore this line. However, there is no other solution than the $Z$ point in the current situation. So we must keep this point as the solution. But after this choice  of the angles $\theta_f$ and $\varphi_f$, we will have the trivial situation with the zero initial entropy. So the current measurement is the last measurement that can be realised and after it the collapse is impossible (see the comment after Eq. (\ref{Ax1})). 

These two examples have clarified a general feature of the information model of the collapse phenomena. On the whole, if the initial entropy is not zero then the collapse can be described. If we recurrently try to apply the model to some state (`recurrently' means that the final state and angles after the first measurement are the initial state and angles for the second one and so on), because of the second axiom we can reduce the initial entropy to zero, afterwards there is no collapse. We will call this reduction as the death point effect.

\section{The model of Everett's worlds}

In the previous two sections, the first two axioms have been discussed. In this section we are going to dispute the last axiom and try to build the function of risk $\R$. 

In our problem: the collapse of a wavefunction at the measurement of a spin $1/2$ projection, instead of the $\R$ function which needs minimizing, we can use more simple approach. First of all, let us consider the boolean set $\{0,1\}$. Hereinafter, we will interpret $1$ as the $+1/2$ projection and $0$ as the $-1/2$ projection  
$$ 1 \equiv \Up, \qquad 0 \equiv \Dw.$$
{\it Using this notation we can introduce a boolean function $\p$, which will be employed instead of the $\R$ function, and its value will represent the final projection after the collapse.} As it can be seen, later this approach is more natural than the previous one. The function of risk is useful when we have an infinite set of the final state, for example, in the case of week measurements.

Supposing the $\p$ function is a function of $\theta_f$ and $\varphi_f$ $\p=\p(\theta_f, \varphi_f)$ (this proposal will be generalized later), we reduce the problem of the construction of the form of the $\p$ function to the problem of the building of a boolean function of continuous variables.

The simplest method how to achieve this is the following. Let us introduce boolean variables 
$$ \xi = \sigma(\cos\theta_f), \qquad \eta = \sigma(\cos\varphi_f), \qquad \sigma(x) = \left\{ \begin{array}{l} 1 : x>0 \\ 0 : x\le 0 \end{array} \right. .$$
The $\sin(x)$ function cannot be applied here, because it is always positive on $(0, \pi)$. The manifest form of the $\p$ function can be determined by fixing its truth table. Knowing this table, we are able to represent the function by disjunctive normal form or conjunctive normal form \cite{Schneeweiss}.

Following this idea, we can construct the most general form of the $\p$ function. Let us consider a series of $n$ consequence collapses. By $\theta_f^i$ and $\varphi_f^i (i=1,\cdots, n)$ we denote the final angles after $i$ collapse, and $ \xi_i = \sigma(\cos\theta_f^i)$ and $\eta_i = \sigma(\cos\varphi_f^i)$ are their boolean projections; $s_i$ is the spin projection after $i$ collapse in boolean representation. In the most general case, the $\p$ function can depend on these boolean parameters as well as on $\xi$ and $\eta$ which represent the final angles in the current measurement, i. e. 
$$ \p = \p \Big( \xi_1, \eta_1, s_1, \, \xi_2, \eta_2, s_2, \, \cdots, \, \xi_n, \eta_n, s_n; \, \xi, \eta \Big). $$

If $n\neq0$ then we include the effect of memory in our model, i. e. the result of the collapse at the current measurement depends on the results of the $n$ previous measurements. It is obvious that the properties of the $\p$ function are fixed by $n$. 

Let us employ the concept of the $\p$ function to Everett's interpretation of quantum mechanics. According to Everett's concept, different terms of a quantum mechanical superposition correspond to different classical worlds (see for example \cite{Menskii, Mensky}). In the proposed model of the collapse phenomena, the $\p$ function determines the projection of a spin after a measurement, i. e. this function defines in what classical world the observers will be after the collapse. Therefore, {\it a classical world corresponds to the set of the $\p$ functions that lead to this world after the collapse.} Moreover, in our model, transitions between classical worlds, that have been mentioned in \cite{Menskii, Mensky}, can be easily described like {\it the replacing of the $\p$ function.}

It ought to be mentioned that the information model of the collapse phenomena is closely related to the abstract mathematical structure under the name {\it `automaton'} (see for example \cite{Gecseg}). In order to illustrate this connection, it is convenient to quote the definition of an automaton here.
\begin{quotation}
``A {\it Mealy-type automaton} [...] is a system $\mathbf{A}=(A,X,Y,\delta,\lambda)$ where $A,X$ and $Y$ are (non-empty) sets; furthermore, $\delta : A \times X\to A$ and $\lambda: A\times X\to Y$ are functions defined on $A\times X$. Sets $A, X$ and $Y$ are the {\it sets of (internal) states, inputs} and {\it outputs}, respectively. Functions $\delta$ and $\lambda$ are called {\it transition function} (or {\it next state function}) and {\it output function}, respectively.''  (\cite{Gecseg}, P.~1)
\end{quotation}
In our case, the set of states $A$ is the set of the states of the observer, i. e. the set of all $\sigma(\theta,\varphi)$; $X=Y$ and equal to the Hilbert space of all spin projections; the first and second axioms realise the transition function $\delta$; the $\p$ function is the output function $\lambda$. Now it can be easily seen that using the $\p$ function approach naturally leads to the automaton formulation of the information model of the collapse. Moreover, taking into account the Extended Everett Concept \cite{Menskii, Mensky}, this formulation allows us to build the mathematical model of consciousness: {\it consciousness can be modeled by the automaton that has been mentioned above.}

In the end of this section, we are going to illustrate the $\p$ function approach in two cases of the simplest $\p$ functions: 
$$ \p_{\vee} = \xi\vee\eta, \qquad \p_{\wedge} = \xi\wedge\eta,$$
where $\vee$ is the disjunction or OR-function and $\wedge$ is the conjunction or AND-function \cite{Schneeweiss}.
\begin{figure}
\begin{center}
\includegraphics[scale=0.2]{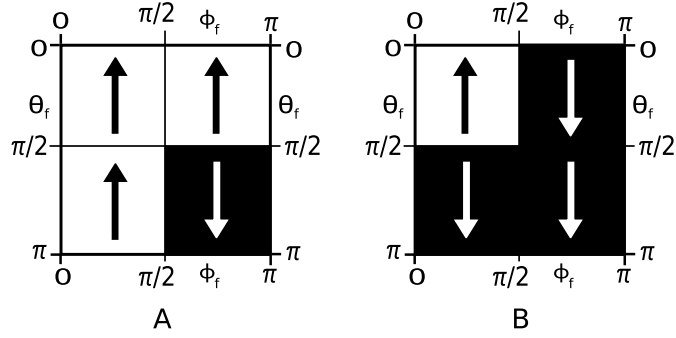}
\end{center}
\caption{The simplest $\p$ functions}
\label{fig5}
A -- the plot of the $\p_{\vee}$ function, B -- the plot of the $\p_{\wedge}$  function. The region of the black color represents the region of the spin down, the region of the white color -- the spin up.  
\end{figure}
The plots of these functions are presented on Fig. \ref{fig5}. From these plots, the following important property of the $\p$ function approach can be seen. If the probability of obtaining the spin up projection  as well as the spin down projection can be treated as a geometrical probability then they are not equal. From the point of view of the information model of the collapse, the last fact confirms the following assumption that ``{\it consciousness may modify probabilities of classical alternatives}'' (\cite{Mensky}, P.~8).

\section{Conclusions}

We have further advanced the information model of the collapse phenomena. The $\p$ function approach has been proposed to construct the manifest form of the function of risk. We have exposed the death point effect, as an important property of the model, when we tried to carry out numerical calculations. A number of joints with the Extended Everett Concept \cite{Menskii, Mensky} have been detected.

The formulation of the model in terms of the automaton ought to be the object of future investigations, because this formulation unites the three axioms in one mathematical structure. The second interesting aim of the studies is the death point effect. According to the death point effect, if an observer has reached the point of zero entropy, after this he has a trivial situation when he wants to measure the same projection of a spin that the electron has. Therefore, hereinafter there is no collapse and the observer does not act as if he is {\it `dead.'} According to \cite{Menskii, Mensky}, the observer is able to transit to another classical world. From this aspect, the death point effect might be the reason of these transitions. In other words, wishing to escape from `his death', the observer does need changing his current classical world. The third important point that needs further investigations is to offer an experiment for the inspection of our model.
\begin{acknowledgments}
The author wants to thank Dr. Robert Lompay for very important discussions. 
\end{acknowledgments}

\bibliography{Applications}

\end{document}